# Blockchain Based Secured E-voting by Using the Assistance of Smart Contract

Kazi Sadia[1[0000-0002-2841-054]], Md. Masuduzzaman[2[0000-0001-8039-367]], Rajib Kumar Paul [3[0000-0002-6191-832]], and Anik Islam[4[0000-0002-6725-9805]]

[1,2,3] Department of Computer Science and Engineering, American International University-Bangladesh, Dhaka, Bangladesh
{qazirisha, masuduzzaman.prince, krajib60}@gmail.com
[4] Department of IT Convergence Engineering, Kumoh National Institute of Technology, Gumi, 39177, South Korea
anik.islam@kumoh.ac.kr

**Abstract.** Voting is a very important issue which can be beneficial in term of choosing the right leader in an election. A good leader can bring prosperity to a country and also can lead the country in the right direction every time. However, elections are surrounds with ballot forgery, coercion and multiple voting issues. Moreover, while giving votes, a person has to wait in a long queue and it is a very time-consuming process. Blockchain is a distributed database in which data are shared with the participant of the node and each participant holds the same copy of the data. Blockchain has properties like distributed, pseudonymity, data integrity etc. In the paper, a fully decentralized e-voting system based on blockchain technology is proposed. This protocol utilizes smart contract into the e-voting system to deal with security issues, accuracy and voters' privacy during the vote. The protocol results in a transparent, non-editable and independently verifiable procedure that discards all the intended fraudulent activities occurring during the election process by removing the least participation of the third party and enabling voters' right during the election. Both transparency and coercion are obtained at the same time.

**Keywords:** Blockchain, E-voting, Hash, Security, Smart contract.

## 1 Introduction

### 1.1 Blockchain

Blockchain is essentially an open, distributed database of records or a public ledger of all transactions or digital events that have been occurred and shared among participating parties connected within a network [1]. A blockchain is basically a chain of blocks where blocks are connected to form a chain of blocks that holds data or information regarding any event [2]. Each transaction or activity within the blockchain is verified by consensus of a majority of the participants i.e. without the approval of the majority network, an activity cannot be taken into consideration [3]. Once some data has been inserted into a blockchain, it becomes very difficult to change it due to having immutability configuration [4]. In order to re-write any data, dishonest miners must re-write the previously broadcasted block, and the changes have to be agreed by the other miners in the network [1].

In blockchain, double spending is prevented by the computational power “Proof of work” that requires computer processing power to generate fingerprints to uniquely identify each block [1]. Blockchain technology uses cryptography which ensures the legitimacy of a transaction [5]. Third party involvement is prevented by the peer to peer network validation therefore, cost and trust related issues are resolved [6]. The structure of a block in blockchain is described below.

**Data**- The data can be the type of information is stored in the block.

**Hash**- The hash is a kind of fingerprint that uniquely identifies a block and is determined by its contents.

**Hash of previous block**- Points the previous block to form the chain, a change in a single hash causes the after created blocks to change their hash.

According to Fig. 1, when a participant intends to add a block to the chain, the peer nodes are responsible for validating the. After the verification, the majority agrees to add the block and thus, the block is added to the blockchain [7]. If majority denies, the block is discarded.

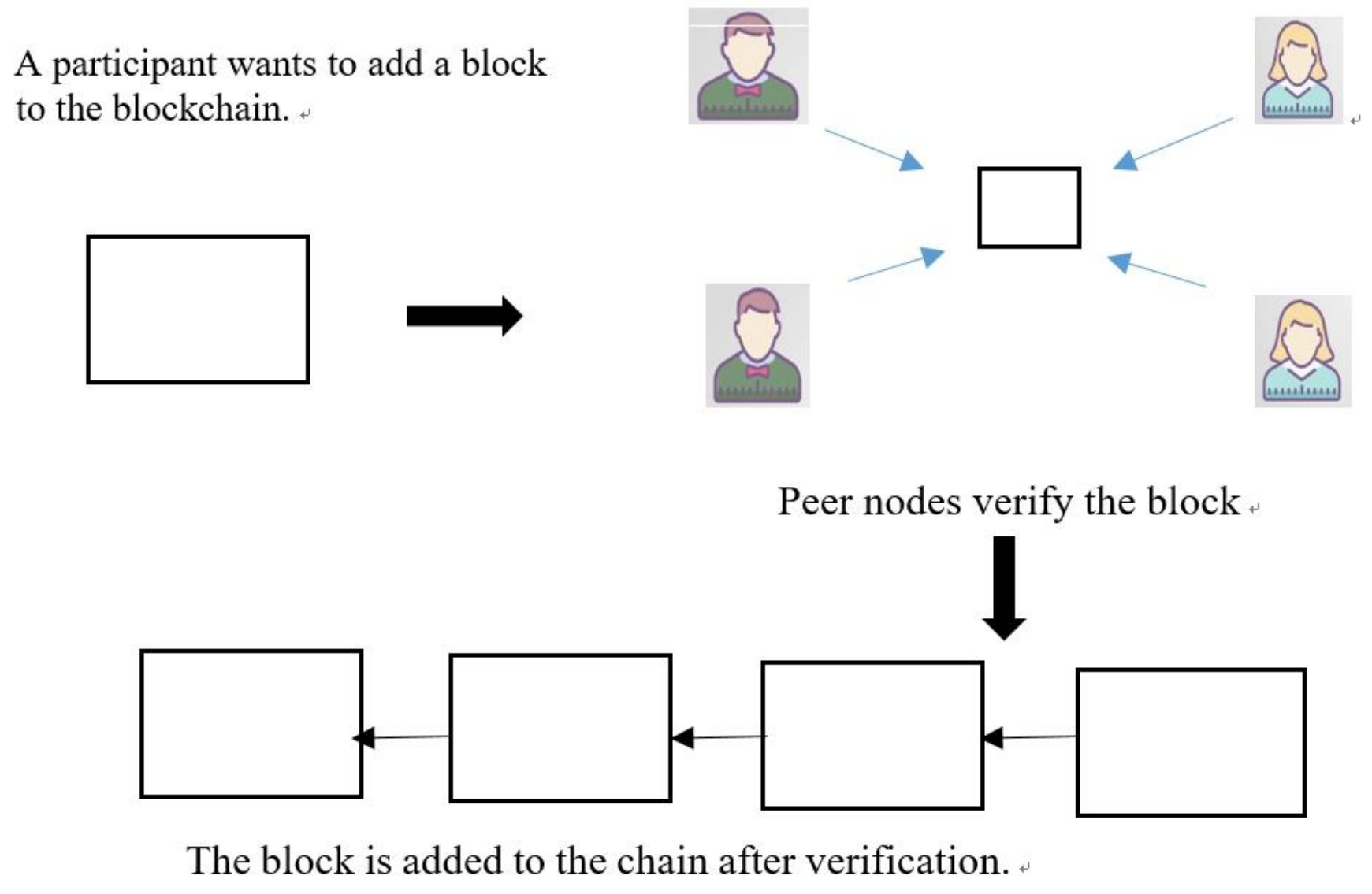

**Fig. 1.** The mechanisms of adding a block in blockchain

## 1.2 E-voting

In every democracy, the main important thing is to secure the election process for the national security and development of a nation. Ever since the candidates were needed to be elected through a democratic process, it was done by voting with pen and paper. Afterwards the result was counted manually and declared. The process of voting with paper ballot and pen required a lot of time and created hustle in maintaining long queue. Also, the manual process ensued in ballot forgery, coercion and multiple voting. Now replacing the bygone process of electing with pen and paper by a new innovative process might be condemnatory in stopping any sort of duplicity and forgery which is traceable and consequent [8].

E-voting is the new concept proposed to ensure fair and digitalized voting that promises to resolve all the issues related to manual voting. By electro voting, we generally mean the vote casting process with the help of any sort of computers or computerized voting equipment or the internet. The tasks are conducted through systems to hereby reduce the involvement of manpower during the election process. Registering the voters, tally ballots and recording of votes can also be easily done by this electronic system [9].

Electro voting machine neither a complex machine nor harder to operate. It can be easily understood and operated by both election officer's in-charge and the voters. EVM has basically two units- Control unit, Display unit and Ballot unit. The main unit of EVM is control unit which stores all the data and controls the basic function including voter information. Vote counting is assuredly conducted with possibly less time and accuracy.

## 1.3 Smart Contract

Previously, contracts between parties were held upon visual meetings, often fully self-executing or self-enforcing. The Smart Contract is aimed to provide contracts between parties where both parties are given the priority and contracts are conducted upon establishing the conditions of both the parties [4]. It is the executable code that runs on top of the blockchain to facilitate the terms required in an agreement of a contract between the two parties. The involvement of any third party is resolved as any medium between parties are not required, upon fulfilling conditions contracts are self-executed.

Smart contract is a legal application that runs on a blockchain network [4]. Smart contracts are much like legal contract. Smart contract can be used in many different things. Banks for example could use it to issue loan, worth for automatic payments, both the e-commerce and music rights managements can use this. Insurance company could use it for process claims, postal company use it for payment on delivery and so on. Smart Contract is like this:

- No trust issue in smart contract just like this vendor machine, as shown in Fig. 2. A person itself can put the coin into this and get the desired product.
- No involvement of third party. The same as this vendor machine. When a person itself involved with this matter can directly interact with it and get the desire product. And there is not any involvement of third-party.

- As the smart contract is distributed in open ledger. There is no chance of lose or hacking. As like in an open environment it is difficult in involve in and manage to steal stuff.

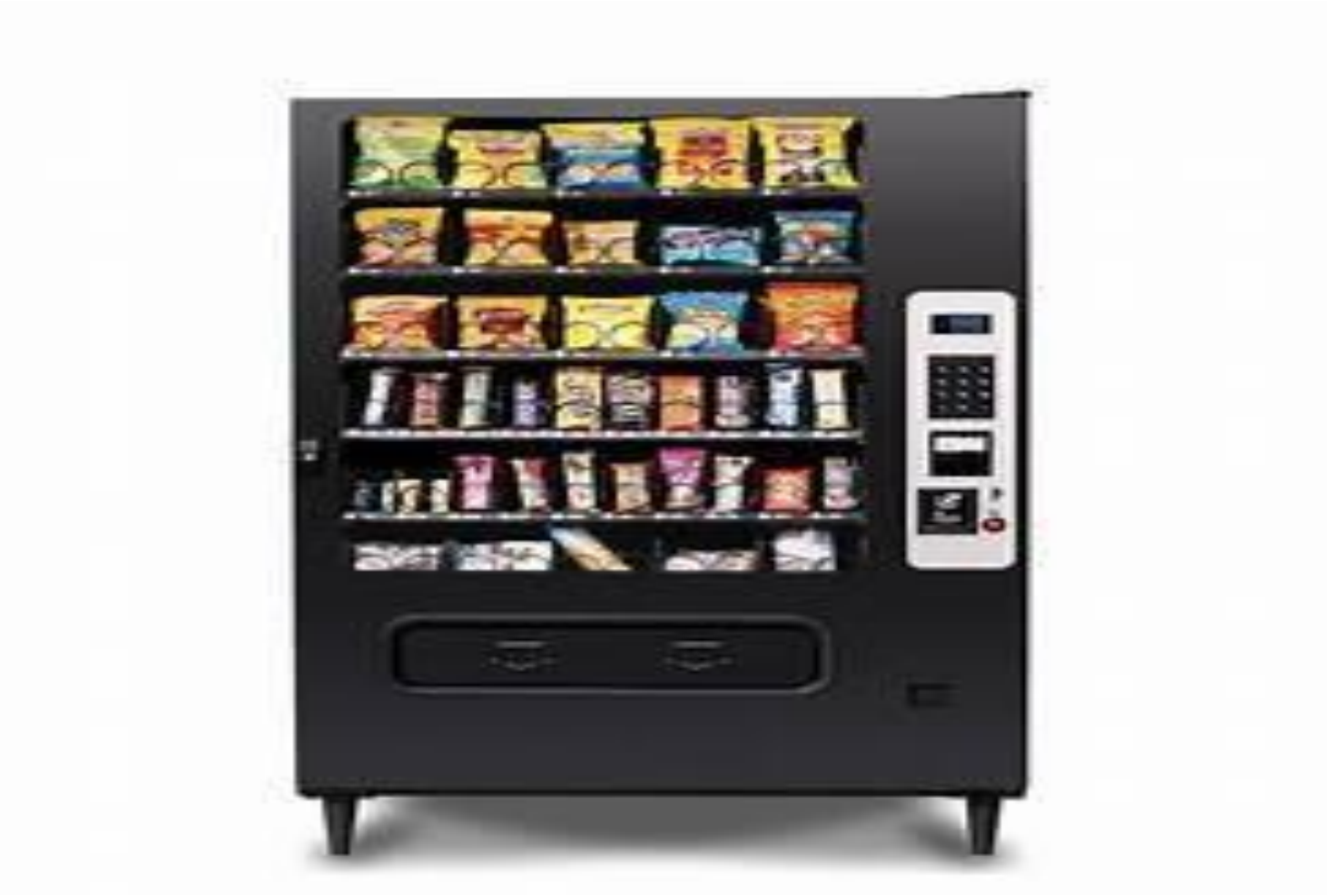

**Fig. 2.** Example of Smart Contract.

The remaining sections of this paper are organized as follows: Section 2 represents related works. In Section 3, proposed scheme is depicted. A security analysis based on different properties is outlined in Section 4. Finally, Section 5 draws a conclusion from this paper.

## 2 Related Work

There has been a lot of work on blockchain based e-voting using cryptography, other signatures and techniques. In such papers, minimal involvement of third party is observed and a problem of coercion and transparency maintenance at the same time is also observed. In fact, the balancing of transparency and coercion-resistance was a possible future work in [10]. Reduction of third party is a major portion of work in an election process, the impact of third-party involvement can have a vulnerable effect on the whole procedure. Moreover, coercion-resistance is a difficult task that is to be mapped with transparency.

Antony Lewis et al. [1] described blockchain as an open, distributed ledger of historical records that uses cryptography and digital signatures. In his paper he also mentioned the logic of blockchain, how does it work. On explaining the aftermath of resolving conflicts, he introduced an idea of not broadcasting a block intentionally. Two blocks can be created, and one can be left as being not broadcasted, the un-broadcasted block can be broadcasted when desired. In this paper, we have used this concept to keep the choices of nominee secured until result calculation.

Y. Liu et al. [10] proposed a protocol where the choice was made safe using random string and choice code. The length of the vote string varies depending on election requirements. The choice code represented the voter's choice followed by random string which is an indication of well-formed vote. According to Y. Liu et al. [10], there are Pre-voting phase, Voting Phase, Post Voting Phase. In the Pre-voting phase, the organizer Bob collected all valid ballots. After ending the voting time, Bob generates a set of all ballots which means that all the ballots has been received. Then Bob runs this algorithm 1:

**Algorithm 1** To Obtain All Valid Ballots

**Input:** *AllBallots*: the set of all ballots Bob has received
**Output:** *ValidBallots*: the set of all valid ballots
1: **for each** $b \in Ballots$ **do**
2: **if** $isCorrectFormat(b)$ & $hasAllSignature(b)$ & $isCastOnTime(b)$ & $hasNotBeenCounted(b)$ **then**
3: $ValidBallots \leftarrow ValidBallots \cup \{b\}$
4: **end if**
5: **end for**

This algorithm runs to gain set of Valid Ballots which is set of all the valid ballots.

There are issues regarding an election, so voters' privacy must be assured thus concept of public and private keys upon reviewing different papers are used but with a little modification. Anonymity was ensured by keeping voters' identity private [10][11]. According to Y. Liu et al. [10] and Freya Sheer Hardwick et al. [11], one must authenticate oneself to the Central Authority and CA receives a token that proves one's eligibility to vote. In these papers, one central authority or an officer is responsible for initial verification.

The counting phase described in the protocol discussed by Freya Sheer Hardwick et al. [11], deals with broadcasting a ballot opening message which contains value that will represent the voter's choice and the voter's themselves broadcast this. Freya Sheer Hardwick et al. [11] stored the information of list of candidates and voters in the genesis block as the initial storage. The authors revealed the result at the end of election using the concept of value representation of the voter choice. A voter can vote multiple time and every time the previous vote was replaced by the current one. By this process coercion is said to be totally removed. In both the papers, everyone can view the public blockchain and there is no centralized authority. In [12-17], they also proposed voting mechanism which utilizes blockchain.

# 3 Proposed Methodology

## 3.1 Procedure

The basic functionalities of the proposed protocol are illustrated accordingly in Fig. 3. a. The code is executed on the top of the blockchain. Therefore, verifying actions that was supposed to be performed by the third party. Moreover, the peer network connected are in charge of further verification as mentioned. The figure introduces some unknown terms that are further described below.

Condition 1 - Verify whether the voter is in the group X and the flag of X is true. Also, check whether the voter is in the eligibility list or not.

Condition 2 - Mathematical computation (proof of work) is done. Also, verify whether the voter has casted vote previously or not and check the ballot is in correct format or not.

Organizer - In this protocol, organizer is the only representative who is involved within the protocol but for a limited time. The role of the organizer is to arrange and collect the list of nominees, list of eligible voters, start date and time, end date and time. The start and end (date and time) are decided and announced by the election commission. The list of eligible voters is collected through manual registration.

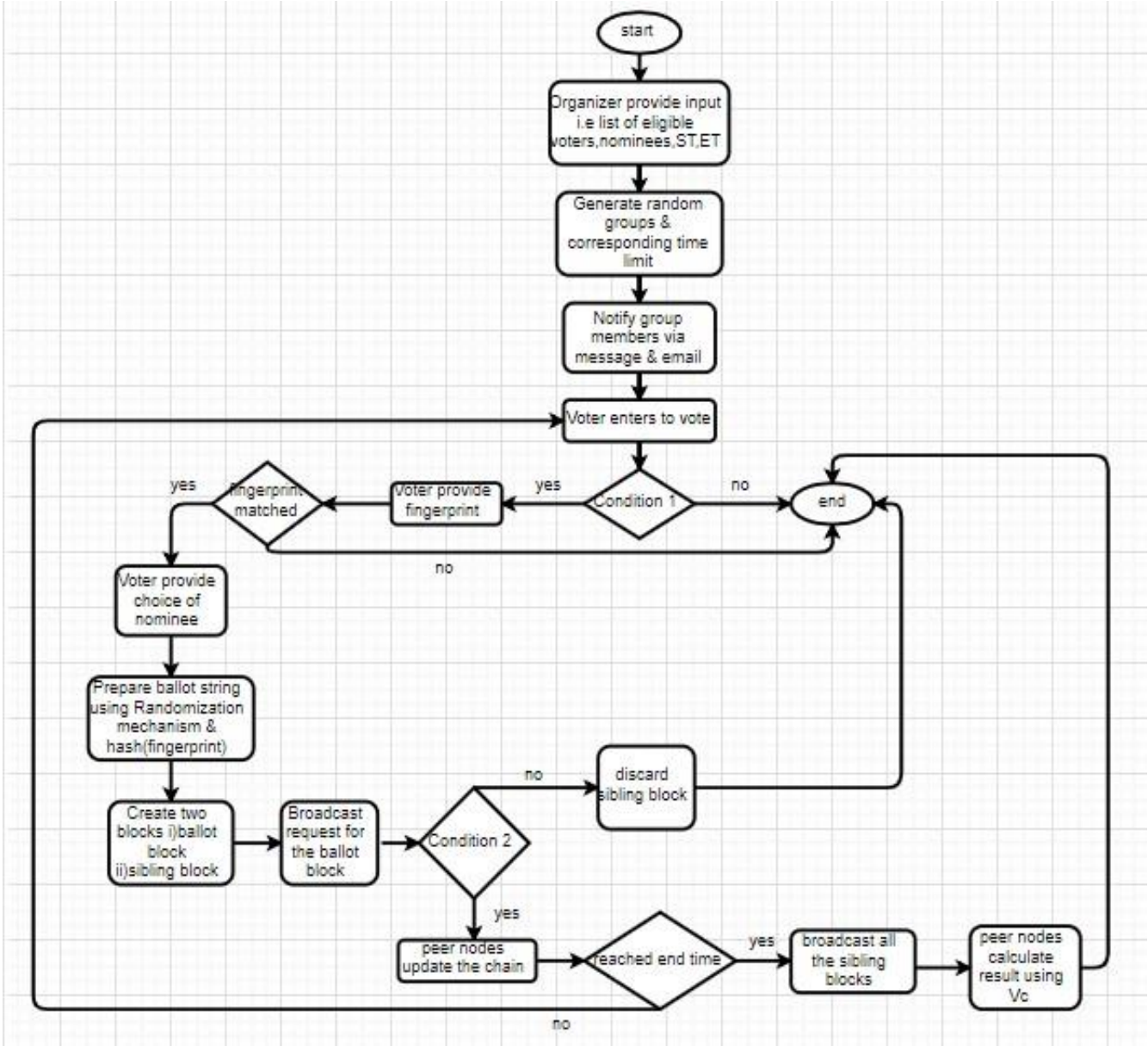

**Fig. 3.** Flowchart representation of the basic functionalities of the protocol.

Ballot string - The string that contains the choice of nominee hidden around random numbers to avoid recognition.

Sibling block - A block that contains the arrangement of choice value.

st - ST - start time

et - ET - end time

Hash(fingerprint) - hash function used on the binary value of the voter's fingerprint

**NOTE:** Choice of nominee is hidden in the ballot string. The arrangement of the choice is hidden in the variable $V_c$, the arrangement is prepared by random number generation. Thus, nobody has any idea of the voter's choice until the end of the election.

### 3.2 The phases of the proposed protocol

The protocol is categorized into three phases, in which each phase is dependent upon another. Following are the three phases:

1. Pre-voting phase.
2. Voting phase.
3. Post voting phase.

**1. Pre-voting phase:**

The organizer is the actor responsible for collecting the list of the eligible voters and nominee based on the desired condition (if any). The list of the voters should contain voters' name, national identification number (NID), fingerprint and any other information based on the direction of the election commission. Organizer provides the list of eligible voters, their fingerprint coordinates along with the binary value, nominees, start date-time and end date-time as an input on the genesis block. In case of people having problem, alternative option is considered. Priority is maintained, thumb is given the first priority, and people deprived of thumb can use the grooming finger. For worst case, message verification process can be used. In that process, a pin code is sent to the particular contact number of the voter, the voter is to provide the pin in order to verify himself as an alternative of fingerprint. Genesis block is the parent block or the first block of the blockchain. The start date-time and end date-time is mentioned earlier by the election commission. The role of the organizer ends here; as per the result of the code execution, the procedure is carried out. The program (code) is previously integrated within the blockchain as per the concept of Smart Contract. On reaching the start date-time, one of the pre-defined condition fulfills i.e. {if (DateTime.Now==st) start ();}; a function is called which invokes the election procedure to start and corresponding activities are performed. Voters are grouped randomly based on the number of eligible voters and any other condition provided, and random time is generated for each group. Each group will have distinct timing; overlapping is not taken in consideration. Voters of specific groups are notified via email and message; a time limit is set for each group.

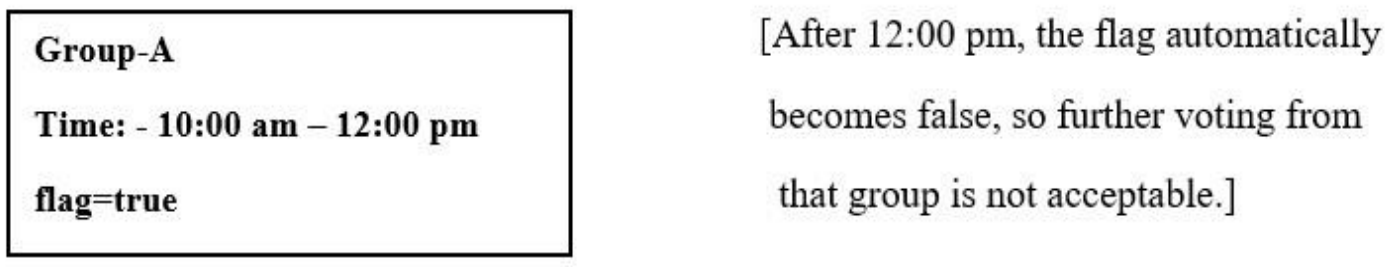

The flag is a Boolean property of a group. The flag remains true till the time limit of the group. The duration of each group is also decided by the election authority. The voting duration for each group must be adjusted in such a way that none of the voters skip to vote due to load/traffic on the network at that instant. The code works upon that to generate different timing for each group. No one is allowed to vote after the flag becomes false i.e. the time limit exceeds. The flag becomes false automatically once all the voters within the group are done with their voting (this provides further security)

## 2. Voting phase:

As the voter approaches to vote providing his/her public keys, it is verified (within the code) whether the voter is in the group with a flag value of true and whether the voter is in the eligibility list. As smart contract performs an executable code, it is verified through the code by the call of a function that checks whether the voter entered is eligible or not. Given that, the eligibility lists of the voters are stored on the genesis block. As the voter has proved him/her as eligible and also the voter is in the specified group (the group to serve currently), the voter is then to provide his/her private key (fingerprint which is converted to binary data, as shown in Fig. 4) as a need of verification that no other people except the voter is casting his/her vote (this reduces the chance of anyone knowing one's public keys and using the public key to cast vote in the name of the voter, stealing votes). This is the second phase of verification of voters. The fingerprint is matched with the one provided along with the eligibility list in the genesis block. The fingerprint sensor is used to figure out the coordinates of particular voters, the coordinates are then matched with the coordinates provided in the genesis block, if it matches then according to the Fig. 4, the binary value of the coordinates is obtained from the provided list in the genesis block. Conversion of the coordinates into the binary value during the voting process will require time and memory consumption, thus this procedure is performed. The hash of the binary value is basically the unique voter identification in the ballot within the block. Direct voter's identity is avoided to ensure voter's security. The hash $(\mathrm{fingerprint}_{\mathrm{binary}})$ value is the representation of the voter in the block. SHA-256 is used as the secured hash function, hash $(\mathrm{fingerprint}_{\mathrm{binary}})$ that cannot be reversed. According to some research, fingerprint is one of the most secure metadata of a person, thus fingerprint is used instead of any other metadata in this protocol.

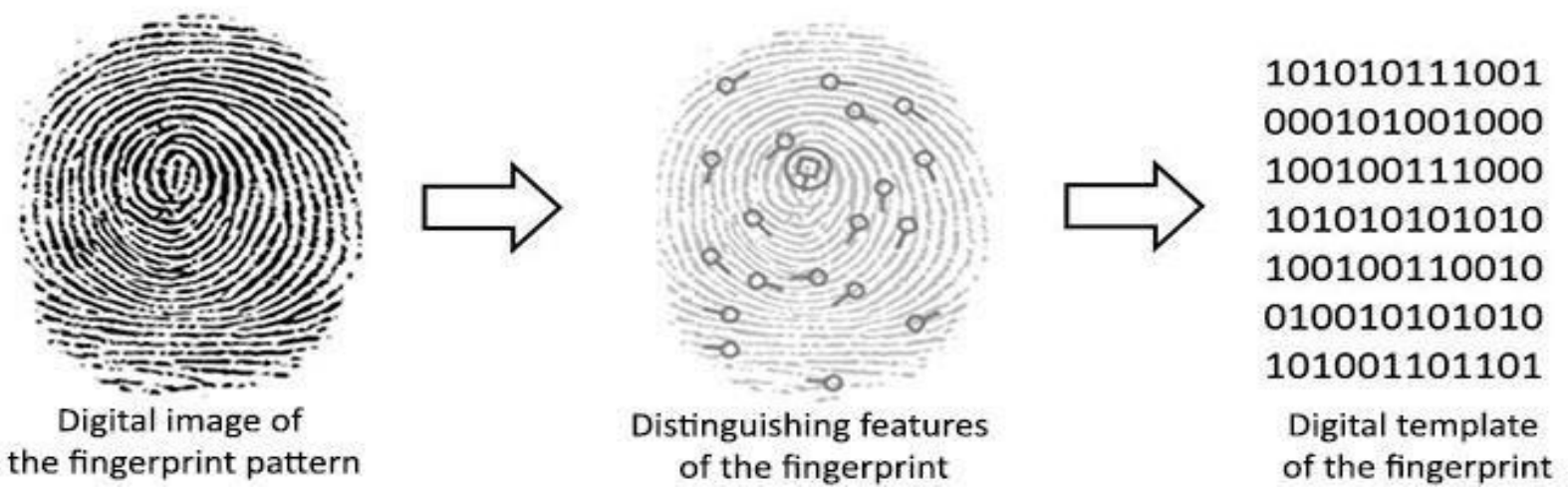

**Fig. 4.** Conversion of the fingerprint pattern to binary value.

The voter is then provided with the list of nominees each represented by a logo. The voter then selects his/her choice of nominee. The nominees are represented by their representative logo, the logos have a binary value which is basically selected and worked with when chosen. The calculations and workings are done upon the distinct binary values. Fig. 5 shows an example of the representation. The number of 1's and 0's in the value representing the nominees must be same otherwise it is possible to guess the choice of nominee in the ballot string. Upon several workings, it is seen that unequal number of 0's and 1's for every nominee may result in prediction of the selection of nominee, as a result the progress of the election is made visible. If the representation does not remain consistent or if it is not possible to allocate different representation of nominee within (N) bits, then increase the number of bits in order to get different representation for equal numbers of 0's and 1's. Example- Three bits with two 1's and a 0 will have representations – 110,011,101 so three logos can be represented by these in other words three nominees can be represented.

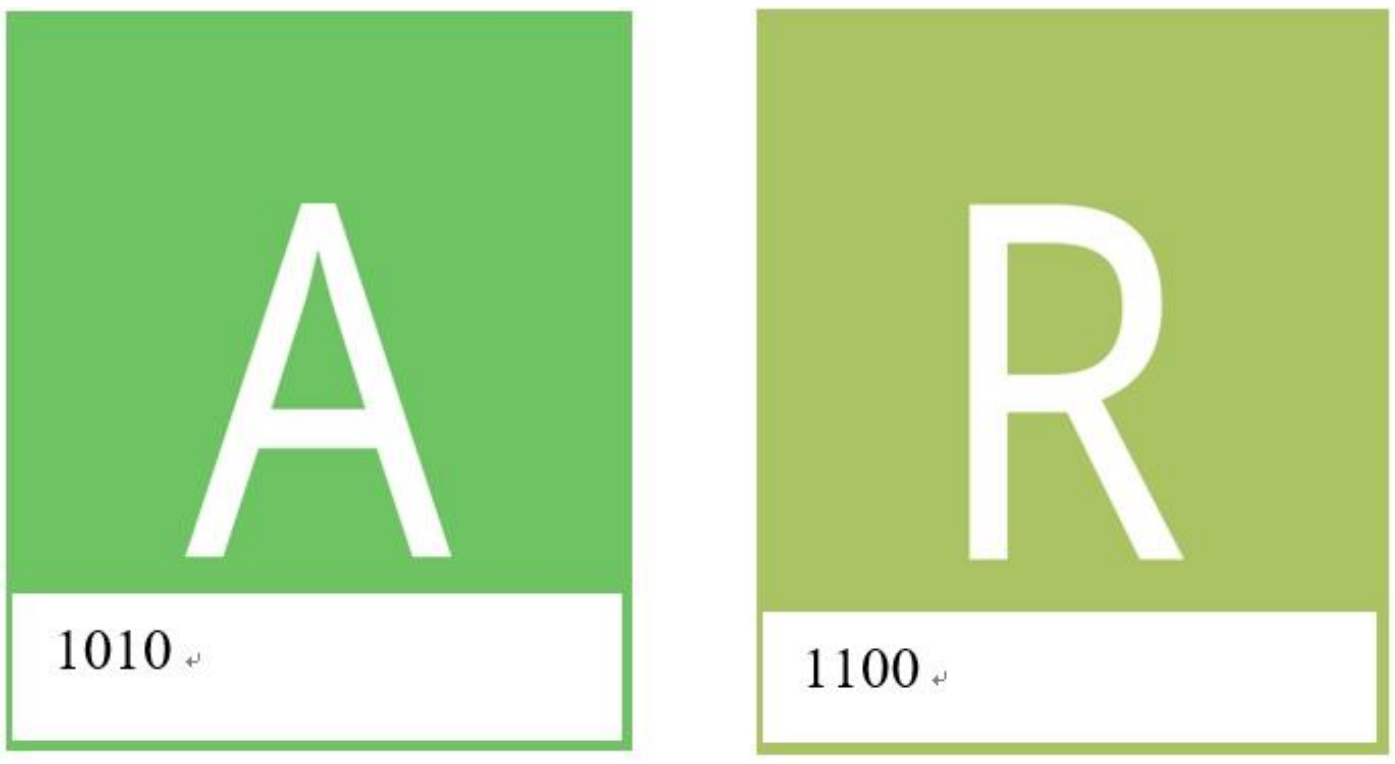

**Fig. 5.** Binary representation of nominee logos.

On choosing the nominee, the preparation of ballot takes place. A ballot is designed to have a ballot number in it in order to refer it. The ballot consists of the voter s' hash ($\text{fingerprint}_{\text{binary}}$) and the ballot string. The ballot string is prepared by the execution of a function inside the code with the concept of smart contract. The ballot string must be different for every voter. The ballot string has two substrings, choice string and the random string. The choice string consists of the nominee choice hidden within other randomly generated values. The random string is randomly generated 0/1 values. These techniques are used in order to prevent viewers from recognizing the choice of nominee. A nominee might get multiple votes therefore, to distinguish every ballot strings the concept of random string is used. Generation of the random string results unique ballot string formation. The ballot string is prepared in two phases, the following are:

**NOTE:** The total number of bits has no restriction. 16-bit is just an example. Greater number of bits is more secure as chances of similar generation of random number decreases. The decision of the number of bits must be taken in consideration before taking decision.

Consider a 16-bit ballot string of which 8 bits are choice string-the red ones and 8 bits are random string- the black ones. The ballot string is equally divided in these two parts.

| | | | | | | | | | | | | | | | |
|---|---|---|---|---|---|---|---|---|---|---|---|---|---|---|---|
| | | | | | | | | | | | | | | | |
| 0 | 1 | 2 | 3 | 4 | 5 | 6 | 7 | 8 | 9 | 10 | 11 | 12 | 13 | 14 | 15 |

i) If n bits are representing each logo then $n$ random numbers are generated from 0-7 as the choice string is between 0-7, the binary value of the logo is arranged in the generated random value indexes of the ballot string i.e. Alice chooses the nominee with a binary value of 1100, the binary value consists of four bits thus four random numbers are generated in order to hide the choice of Alice.

Number of bits representing each logo - 4

Random numbers - 4,5,7,0 (4) – $V_c$ - opening value

Nominee choice - binary value of logo - 1100

Therefore, the four randomly generated number- 4,5,7,1 are the indexes to hide the binary value of the nominee's choice. The value is assigned sequentially.

| | | | | | | | | | | | | | | | |
|---|---|---|---|---|---|---|---|---|---|---|---|---|---|---|---|
| 0 | | | | 1 | 1 | | | 0 | | | | | | | |
| 0 | 1 | 2 | 3 | 4 | 5 | 6 | 7 | 8 | 9 | 10 | 11 | 12 | 13 | 14 | 15 |

ii) Generate another number between 1/0. Fill that number in the other four indexes. Example-1

| | | | | | | | | | | | | | | | |
|---|---|---|---|---|---|---|---|---|---|---|---|---|---|---|---|
| 0 | 1 | 1 | 1 | 1 | 1 | 1 | 1 | 0 | | | | | | | |
| 0 | 1 | 2 | 3 | 4 | 5 | 6 | 7 | 8 | 9 | 10 | 11 | 12 | 13 | 14 | 15 |

The other indexes of the choice string are assigned with either 1/0, but all the other indexes of the choice string must have the same value to avoid recognition of the choice.

iii) Generate random numbers randomly from 1/0 and put on the indexes (8-15) suppose, 11001010

| | | | | | | | | | | | | | | | |
|---|---|---|---|---|---|---|---|---|---|---|---|---|---|---|---|
| 0 | 1 | 1 | 1 | 1 | 1 | 1 | 1 | 0 | 1 | 1 | 0 | 0 | 1 | 0 | 1 |
| 0 | 1 | 2 | 3 | 4 | 5 | 6 | 7 | 8 | 9 | 10 | 11 | 12 | 13 | 14 | 15 |

As 8-15 is the random string part of the ballot string thus eight random numbers either 1 or 0 are generated and assigned sequentially in order to distinguish each ballot string.

The ballot string is prepared, and the choice is hidden inside the string. The choice is recognized by the $V_c$ only. Dispose of the $V_c$ can only result in the consideration of the vote. One block is created containing the ballot and another sibling block is created that consists of the voters' hash ($fingerprint_{binary}$), the reference number of the broadcasted block, its own reference number and the opening value of the choice, in this case- 4,5,7,0. The figure below shows the arrangement.

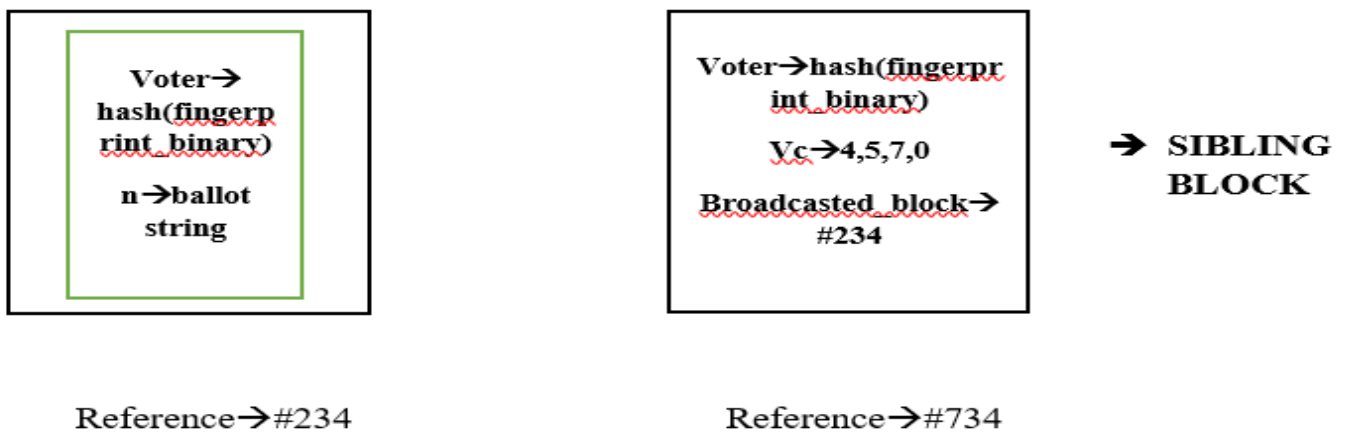

**Fig. 3.** Blocks.

As the voter casts the vote i.e. the voter broadcast the ballot containing block. The ballot contained block is requested to add in the chain whereas the sibling block is remained un-broadcasted. The peer nodes start to work for the proof of work for the block. The one completing earlier then also verifies whether the voter has casted vote earlier and whether the ballot is in correct format. After all the verification, the ballot contained block is added in the blockchain and other peer nodes verifies and updates their chain. Majority is taken in consideration. If majority disagrees then the block is discarded. This not broadcasting issue is pre-defined earlier in the code to ensure the result is calculated only after the election ends.

**3. Post voting phase:**

Once the ending time is reached, it is checked whether all the voters have voted or not. If not, then they are shortly given a notification to complete their voting within specified time. If they fail to do so then no consideration is taken to be granted, otherwise their voting is performed similarly as per the voting phase. If all the voters are done with their voting then as per the contract and execution of codes, all the sibling blocks are broadcasted one by one sequentially. Once all the sibling blocks are broadcasted, the peer nodes start to calculate the result referencing the blocks and using the $V_c$ to extract the choice of nominee for every block. Here all the nodes are supposed to come up with the same result as no blocks are discarded unnecessarily in between and the blockchain supports no changes. Therefore, the voters in other words the peer nodes themselves count the votes and broadcast the result preventing the need of the counting using third party. The blockchain is transparent and the accuracy is ensured as everything is made visible.

## 4 Security Analysis

In blockchain, smart contract and e-voting, security is the main concern that must be taken into consideration at first. Because if the voters are not assured of their safety, they will not get involved to the protocol. Following are the certain security goals which can be satisfied with our proposed methodology.

### 4.1 Anonymity

This protocol uses public and private keys of a voter during the process execution. On the blockchain, only the voter's public key is broadcasted which is hashed previously. Therefore, excluding the voter, no one will be able to recognize any voters within the blockchain due to containing the hash of fingerprint which is basically the binary values of the coordinates.

### 4.2 Voters' privacy

Voters' themselves are not aware of their timing to vote so therefore the chances of manipulation and coercion by fraudulent supporters is reduced. The timing of the voters' voting is only kept within the randomly generated time against each group in the code executed. As a result, manipulators or party specific public cannot blackmail or threat voters.

### 4.3 Confidentiality

Confidentiality is pretty equivalent to part of privacy. The prevention of the sensitive information from reaching unauthorized users while making sure that the right people is in fact aware of it. Most common method ensuring confidentiality is Data encryption and here in this protocol, the data are basically the voter's identity and the ballot string which is encrypted and can only be made visible to all the participants once the election process is over. Being concerned about the voter's identity, only the voter themselves are aware of their identity and choice. The ballot string is prepared in such a way that without the $V_c$, the choice is not understandable. Once the sibling block is broadcasted, the choices are visible to all the peer nodes within the network.

### 4.4 Ballot manipulation

In this protocol, inappropriate ballots i.e. one voter voting more than once is prohibited by the rejection of the approval of the peer nodes. Upon verifying, the peer nodes reject the ballot and is not further added to the block. Ballots without correct format are also discarded and it is made sure that the sibling block of the rejected block is discarded simultaneously. Ballots are basically contained in blocks so updating or change is not that possible. A single change in a block leads to changing the whole other blocks linked with in within specified. So, that's a tough task.

### 4.5 Transparency

Blockchain is an open, distributed ledger where each transactions and activities under consideration is made transparent for peer verification, validation and visibility. While things are kept visible, this ensures no fraudulent activities to take place secretly. The fairness and accuracy is obtained through blockchain's property of being transparent.

### 4.6 Public verifiability & individual verifiability

Our protocol provides the opportunity to publicly verify activities or the voting process as it is kept transparent by the help of blockchain. Peer nodes or anyone can monitor activities taking place. Moreover, voter's themselves can make sure whether their vote is taken in consideration or not. If the block containing voter's identity is broadcasted, this ensures that the voter's vote will be taken into consideration as meanwhile no changes are possibly made. This term is known as individual verifiability which is satisfied through the protocol.

### 4.7 Auditability

Results are calculated after the ending of the election process and the whole process is auditable as blockchain keeps the record of whole thing. The rejected blocks and ballots can be monitored at later stages to have an idea of how often fraudulent activities were intended. Smart Contract codes cannot be modified since it is permanently written on the blockchain. But addition of a portion is possible, for such cases in order to add up portion a forcing function is used. The forcing function will be adjusted in such a way that without the collaboration of all the party's representatives and their approval, changes upon code cannot be made.

### 4.8 Consistency & Accuracy

All the peer nodes will have the same record and at the end same result will be obtained by all the participants. For every activity, a consensus mechanism is carried out thus consistency is satisfied. Moreover, meanwhile no changes are incurred, and the consensus mechanism makes the protocol accurate.

### 4.9 Non-Repudiation

The process of non-repudiation is that someone cannot deny something. Therefore, the result obtained cannot be claimed as being unfair or of fraudulent activities as every activity is made transparent and verifiable by the majority network. It is not possible to mess with the majority honest network. Activities are performed by the execution of code where there is no possibility of unfair means.

## 5 Conclusions

As discussed previously, E-voting is an emerging concept or solution of voting to carry out activities with accuracy and reliability. Moreover, blockchain in an interesting and attractive technology that provides transparency of data and is a topic of high demand. As the process of election must be handled with care in order to avoid unusual circumstances and occurring, therefore, this protocol might reduce the constraints of manual voting and other E-voting systems based on blockchain that uses least involvement of the third party. Also, the reduction of third party completely is a proof of healthy election which is enabled by using the assistance of Smart Contract. The coercion is also prevented by the concept of random generation of groups using Smart Contract. The techniques used in the protocol is quite simpler and easily understandable and designed to reduce memory and time consumption to make tasks faster. As a result, this protocol fulfils all the previously defined properties of the referred paper along with the prevention of coercion with transparency. The voters can monitor the whole process and their privacy is also maintained to avoid any sort of issues. Furthermore, a replacement of the metadata can be taken in consideration to make this protocol widely used in all areas.